%% file: example_paper.tex
\documentclass{article}

\usepackage{microtype}
\usepackage{graphicx}
\usepackage{subfigure}
\usepackage{booktabs} 
\usepackage{amsmath}
\usepackage{tabularx}
\usepackage{array}
\newcolumntype{Y}{>{\centering\arraybackslash}X}

\newcolumntype{L}[1]{>{\raggedright\let\newline\\\arraybackslash\hspace{0pt}}m{#1}}
\newcolumntype{C}[1]{>{\centering\let\newline\\\arraybackslash\hspace{0pt}}m{#1}}
\newcolumntype{R}[1]{>{\raggedleft\let\newline\\\arraybackslash\hspace{0pt}}m{#1}}

\usepackage{hyperref}

\usepackage[accepted]{icml2019}

\icmltitlerunning{End-to-End Multi-Task Denoising}

\begin{document}

\twocolumn[
\icmltitle{End-to-End Multi-Task Denoising for \\
           Joint SDR and PESQ Optimization}



\icmlsetsymbol{equal}{*}
\begin{icmlauthorlist}
\icmlauthor{Jaeyoung Kim}{to}
\icmlauthor{Mostafa El-Khamy}{to}
\icmlauthor{Jungwon Lee}{to}

\end{icmlauthorlist}
\icmlaffiliation{to}{Samsung Device Solutions Research America, USA}


\icmlcorrespondingauthor{Jaeyoung Kim}{jykim@stanford.kr} 
\icmlcorrespondingauthor{Mostafa El-Khamy}{m\_elkhamy@ieee.org}


\icmlkeywords{SDR, PESQ}

\vskip 0.3in
]



\printAffiliationsAndNotice{}  

\begin{abstract}
Supervised learning based on a deep neural network recently has achieved substantial improvement on speech enhancement. Denoising networks learn mapping from noisy speech to clean one directly, or to a spectrum mask which is the ratio between clean and noisy spectra. In either case, the network is optimized by minimizing mean square error (MSE) between ground-truth labels and time-domain or spectrum output. However, existing schemes have either of two critical issues: spectrum and metric mismatches. The spectrum mismatch is a well known issue that any spectrum modification after short-time Fourier transform (STFT), in general, cannot be fully recovered after inverse short-time Fourier transform (ISTFT). The metric mismatch is that a conventional MSE metric is sub-optimal to maximize our target metrics, signal-to-distortion ratio (SDR) and perceptual evaluation of speech quality (PESQ). This paper presents a new end-to-end denoising framework with the goal of joint SDR and PESQ optimization. First, the network optimization is performed on the time-domain signals after ISTFT to avoid spectrum mismatch. Second, two loss functions which have improved correlations with SDR and PESQ metrics are proposed to minimize metric mismatch. The experimental result showed that the proposed denoising scheme significantly improved both SDR and PESQ performance over the existing methods.  
 


\end{abstract}

\input{introduction}
\input{RelatedWork}
\input{end_to_end}

\input{results}
\input{conclusion}
\bibliographystyle{icml2019}


%
%
%


\end{document}

%% file: introduction.tex
\section{Introduction}
\label{sec:intro}


In recent years, deep neural networks have shown great success in speech enhancement compared with traditional statistical approaches. Statistical enhancement schemes such as MMSE STSA~\cite{ephraim1984speech} and OM-LSA~\cite{ephraim1985speech, cohen2001speech} do not learn any model architecture from data but they provide blind signal estimation based on their predefined speech and noise models. However, their model assumptions, in general, do not match with real-world complex non-stationary noise, which often leads to failure of noise estimation and tracking. On the contrary, a neural network directly learns nonlinear complex mapping from noisy speech to clean one only by referencing data without any prior assumption. With more data, a neural network can learn better underlying mapping.  

Spectrum mask estimation is a popular supervised denoising method that predicts a time-frequency mask to obtain an estimate of clean speech by multiplication with noisy spectrum. There are numerous types of spectrum mask estimation depending on how to define mask labels. For example, authors in~\cite{narayanan2013ideal} proposed ideal binary mask (IBM) as a training label, where it is set to be zero or one depending on the signal to noise ratio (SNR) of a noisy spectrum. Ideal ratio mask (IRM)~\cite{wang2014training} and ideal amplitude mask (IAM)~\cite{erdogan2015phase} provided non-binary soft mask labels to overcome coarse label mapping of IBM. Phase sensitive mask (PSM)~\cite{erdogan2015phase} considers phase spectrum difference between clean and noisy signal in order to correctly maximize signal to noise ratio (SNR).  

Generative models such as generative adversarial networks (GANs) and variational autoencoders (VAEs) suggested an alternative to supervised learning. In speech enhancement GAN (SEGAN)~\cite{pascual2017segan}, a generator network is trained to output a time-domain denoised signal that can fool a discriminator from a true clean signal. TF-SEGAN  \cite{soni2018time} extended SEGAN into time-frequency mask.   \cite{bando2018statistical}, on the other hand, combined a VAE-based speech model with non-negative matrix factorization (NMF) for noise spectrum to show good performance on unseen data.    

However, all the schemes described above suffer from either of two critical issues: metric  and spectrum mismatches. SDR and PESQ are two most widely known metrics for measuring quality of speech signal. The typical mean square error (MSE) criterion popularly used for spectrum mask estimation is not optimal to maximize our target metrics of SDR and PESQ. For example, decreasing mean square error of noisy speech signal often degrades SDR or PESQ due to different weighting on frequency components or non-linear transforms involved in the metric. Furthermore, GAN-based generative models don't have even a specific loss function to minimize. Although they are robust for unseen data, they typically perform much worse for test data with small data mismatch. The spectrum mismatch is a well known issue that any spectrum modification after short-time Fourier transform (STFT), in general, cannot be fully recovered after ISTFT. Therefore, spectrum mask estimation and other alternatives optimized in the spectrum domain always have a potential risk of performance loss.

This paper presents a new end-to-end multi-task denoising scheme with following contributions. First, the proposed framework presents two loss functions:
\begin{itemize}
	\item SDR loss function: instead of typical MSE, SDR metric is used as a loss function. The scale-invariant term in SDR metric is incorporated as a part of training, which provided significant SDR boost.  
	\item PESQ loss function: PESQ metric is redesigned to be usable as a loss function. The two key variables of PESQ, symmetric and asymmetric disturbances are approximated to be optimized during training.      
\end{itemize}
The proposed multi-task denoising scheme combines two loss functions for joint SDR and PESQ optimization.  
Second, a denoising network still predicts a time-frequency mask but the network optimization is performed after ISTFT in order to avoid spectrum mismatch. SDR loss function is naturally calculated from the time-domain reconstructed signal. PESQ loss function needs clean and denoised power spectra as input for disturbance calculation and therefore, the second STFT is applied on the time-domain signal for spectrum consistency. The evaluation result showed that
the proposed framework provided large improvement both on SDR and PESQ.

%% file: RelatedWork.tex
\section{Background and Related Works}
\label{sec:related}

\begin{figure}[ht]
\vskip 0.2in
\begin{center}
\centerline{\includegraphics[width=62mm]{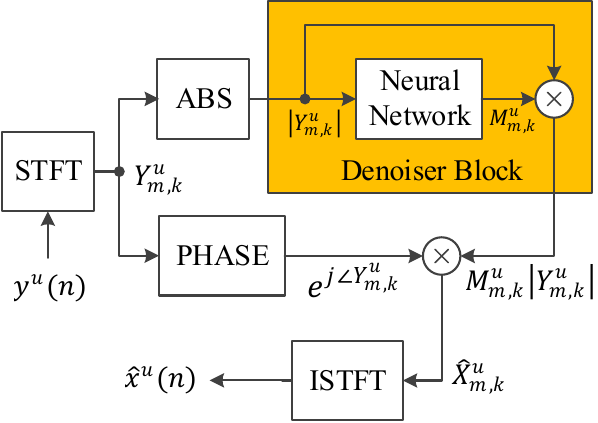}}
\caption{Block Diagram of Denoising Framework}
\label{fig:de_f}
\end{center}
\vskip -0.2in
\end{figure}

Figure~\ref{fig:de_f} illustrates the overall speech denoising flow. The input noisy signal is given by
\begin{equation}
y^u(n)=x^u(n)+n^u(n)
\end{equation}
where $u$ is a utterance index, $n$ is a time index, $x^u (n)$ is clean speech and $n^u (n)$ is noise signal. 
$Y_{m,k}^u$ is STFT output, where $m$ and $k$ are frame and frequency indices, respectively. It is fed into two separate paths. For the upper path, the magnitude of $Y_{m,k}^u$ is passed to the Denoiser block. On the contrary, the phase of $Y_{m,k}^u$ is bypassed without compensation. $\hat{X}_{m,k}^u$ is synthesized by noisy input phase and denoised amplitude spectra. After ISTFT, we can recover time-domain denoised output $\hat{x}^u(n)$.
\subsection{Spectrum Mask Estimation}
\label{sec:sme}
The neural network in Figure~\ref{fig:de_f} predicts a time-frequency mask to obtain an estimate of the clean amplitude spectrum. The estimated mask is multiplied by the input amplitude spectrum as follows:
\begin{equation}
	|\hat{X}_{m,k}^u |=M_{m,k}^u | Y_{m,k}^u |
\end{equation}
Given clean and noisy amplitude spectrum pairs, the mask estimation is to minimize a predefined distortion metric, $d$ for all utterances and time-frequency bins:
\begin{equation}
	\textrm{L} = \min_{\phi} \sum_{u=1}^U \sum_{m=1}^{M_u} \sum_{k=1}^{K} d \left( M_{m,k}^u, | X_{m,k}^u |,| N_{m,k}^u | \right)
\end{equation}	
where $\phi$ is a denoiser parameter, $M_u$ is the number of frames for the utterance $u$, $K$ is the number of frequency bins, $d$ is a distortion metric, $| X_{m,k}^u |$ is an amplitude spectrum of clean speech and $| Y_{m,k}^u |$ is an amplitude spectrum of noisy speech. Ideal binary mask (IBM) and ideal ratio mask (IRM) are two well-known mask labels given by
\begin{equation}
\textrm{L}^u_{\textrm{IBM}, m, k} = \bigg\{
\begin{array}{l}
1,  \textrm{   if   }  \frac{| X_{m,k}^u |}{| N_{m,k}^u |} \geq 10^{\frac{S_{th,k}}{10}} \\
0,  \textrm{   else }
\end{array}
\end{equation}
where $S_{th,k}$ is a log-scale threshold for frequency index $k$.
\begin{equation}
\textrm{L}^u_{\textrm{IRM}, m, k}=\frac{| X_{m,k}^u |}{| X_{m,k}^u |+| N_{m,k}^u |}
\end{equation}
The distortion metric for IBM and IRM is given by
\begin{equation}
d_1\left( M_{m,k}^u,| X_{m,k}^u |,| N_{m,k}^u |\right) = \left(M_{m,k}^u -\textrm{L}_{x,m,k}^u \right)^2
\label{eq:dm1}
\end{equation}
where $x$ is either IBM or IRM. 
The one issue for two mask labels is that they don't correctly recover clean speech. For example, $\textrm{L}_{x,m,k}^u |Y_{m,k}^u |$ is generally not equal to $|X_{m,k}^u |$. 
Ideal amplitude mask (IAM) is defined to represent the exact magnitude ratio:
\begin{equation}
\textrm{L}_{\textrm{IAM},m,k}^u=\frac{|X_{m,k}^u |}{|X_{m,k}^u+N_{m,k}^u |}
\end{equation}
For IAM, instead of direct mask optimization as IBM or IRM at Eq.(\ref{eq:dm1}), \cite{weninger2014discriminatively} suggested magnitude spectrum minimization, which was shown to have significant improvement:
\begin{equation}
\label{eq:dm2}
d_2\left(M_{m,k}^u,|X_{m,k}^u |,|Y_{m,k}^u |\right)=\left(M_{m,k}^u |Y_{m,k}^u |-|X_{m,k}^u |\right)^2
\end{equation}
The one drawback in Eq.(\ref{eq:dm2}) is that it cannot correctly maximize signal to noise ratio (SNR) when phase difference between clean and noisy spectra is not zero. The optimal distortion measure to maximize SNR is to minimize mean square error between complex spectra as follows:
\begin{eqnarray}
\label{eq:dm3}
d_3\left(M_{m,k}^u,|X_{m,k}^u |,|Y_{m,k}^u |,\angle X_{m,k}^u,\angle Y_{m,k}^u \right)= \nonumber && \\  
\left| M_{m,k}^u |Y_{m,k}^u | e^{j\angle Y_{m,k}^u}-|X_{m,k}^u | e^{j\angle X_{m,k}^u}\right|^2  &&
\end{eqnarray}
Eq.(\ref{eq:dm3}) can be rearranged after removing unnecessary terms for optimization:
\begin{eqnarray}
\label{eq:dm31}
d_3\left(M_{m,k}^u,|X_{m,k}^u |,|Y_{m,k}^u |,\angle X_{m,k}^u,\angle Y_{m,k}^u \right)= \nonumber && \\  
\left(M_{m,k}^u |Y_{m,k}^u |-|X_{m,k}^u | \cos(\angle Y_{m,k}^u-\angle X_{m,k}^u) \right)^2  &&
\end{eqnarray} 
The equivalent mask label, also called phase sensitive mask (PSM) label is given by
\begin{equation}
\textrm{L}_{\textrm{PSM},m,k}^u=\frac{|X_{m,k}^u |}{|X_{m,k}^u+N_{m,k}^u |}\cos(\angle Y_{m,k}^u-\angle X_{m,k}^u)
\end{equation}
In this paper, PSM~\cite{erdogan2015phase} was chosen as a baseline denoising scheme because PSM showed the best performance on the target metrics of SDR and PESQ among spectrum mask estimation schemes.  
\subsection{Griffin-Lim Algorithm}
ISTFT operation in Figure~\ref{fig:de_f} consists of IFFT, windowing and overlap addition as follows:
\begin{equation}
\hat{x}^u (n) = \sum_{m=0}^{M_u-1} \hat{x}^u_m(n-m\delta)w_{\textrm{ISTFT}}(n-m\delta)
\end{equation}
where $\delta$ is frame shifting number and $\hat{x}^u_m(n)$ is IFFT of $\hat{X}_{m,k}^u$.
Due to the overlapped frame sequence generation, any arbitrary spectrum modification can cause spectrum mismatch.
For example, $\hat{X}_{m,k}^u$ is, in general, not matched to the STFT of $\hat{x}^u (n)$ due to the modification of amplitude spectrum. Griffin-Lim algorithm~\cite{griffin1984signal} is to find  legitimate $z^u (n)$, which has the closest spectrum to $\hat{X}_{m,k}^u$. A legitimate signal is meant to have no spectrum mismatch. The Griffin-Lim formulation is given by
\begin{eqnarray}
\label{eq:form1}
\min_{Z_{m,k}^u}\sum_{k=0}^{K-1}\sum_{m=0}^{M_u-1}|\hat{X}_{m,k}^u-Z_{m,k}^u|^2 = \nonumber && \\
 \min_{z^u (n)}\sum_{n=0}^{N-1}\sum_{m=0}^{M_u-1}(\hat{x}_m^u (n)- z^u (n)w_{\textrm{STFT}}(n-m\delta) )^2 &&
\end{eqnarray}
where $Z_{m,k}^u$ is STFT of $z^u (n)$ and $w_{\textrm{STFT}}$ is a STFT window function.
The solution can be easily derived by finding $z^u (n)$ that makes Eq.(\ref{eq:form1}) have zero gradient:
\begin{equation}
z^u (n)=\frac{\sum_{m=0}^{M_u-1}w_{\textrm{STFT}}(n-m\delta)\hat{x}^u_m(n-m\delta)}{\sum_{m=0}^{M_u-1}w^2_{\textrm{STFT}}(n-m\delta)}
\end{equation}
Although Griffin-Lim algorithm guarantees to find a legitimate signal with minimum MSE, it is still not guaranteed to coincide with $\hat{X}_{m,k}^u$. The iterative Griffin-Lim algorithm further decreases magnitude spectrum mismatch for every iteration by allowing phase distortion, which will be evaluated in conjunction with the proposed framework.  


%% file: end_to_end.tex
\section{The Proposed Framework}
\label{sec:end_to_end}
\begin{figure}[ht]
\vskip 0.2in
\begin{center}
\centerline{\includegraphics[width=80mm]{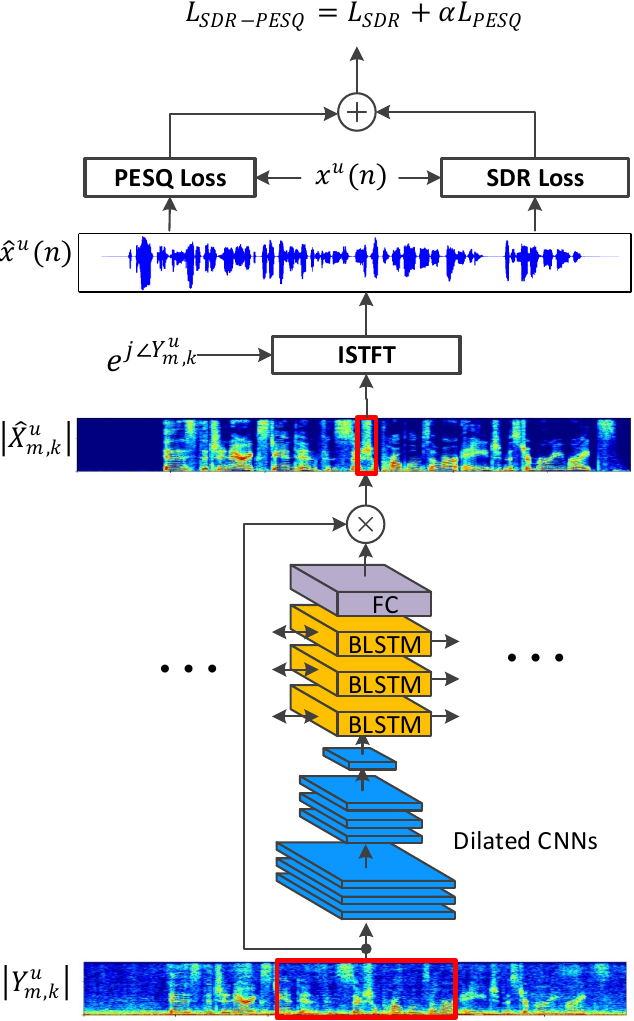}}
\caption{Illustration of End-to-End Multi-Task Denoising based on CNN-BLSTM}
\label{fig:ete}
\end{center}
\vskip -0.2in
\end{figure}
Figure~\ref{fig:ete} describes the proposed end-to-end multi-task denoising framework. The underlying model architecture is composed of convolution layers and bi-directional BLSTMs. The spectrogram formed by 11 frames of the noisy amplitude spectrum $|Y^u_{m,k}|$ is the input to the convolutional layers with the kernel size of 5x5. The dilated convolution with rate of 2 and 4 is applied to the second and third layers, respectively, in order to increase kernel's coverage on frequency bins. Dilation is only applied to the frequency dimension because time correlation will be learned by bi-drectional LSTMs. Griffin-Lim ISTFT is applied to the synthesized complex spectrum $\hat{X}^u_{m,k}$ to obtain time-domain denoised output $\hat{x}^u(n)$. Two proposed loss functions are evaluated based on $\hat{x}^u(n)$ and therefore, they are free from spectrum mismatch.  
\subsection{SDR Loss Function}
\label{sec:si-sdr}
Unlike SNR, the definition of SDR is not unique. There are at least two popularly used SDR definitions. The most well known Vincent's definition~\cite{vincent2006performance} is given by
\begin{equation}
\textrm{SDR}=10\log_{10}\frac{|| x_{target} ||^2}{|| e_{noise} + e_{artif} ||^2}
\label{eq:sdr}
\end{equation}
 There was an $e_{interf}$ term in the original SDR but it is removed because there is no interference error for single source denoising problem.  $x_{target}$ and $e_{noise}$ can be found by projecting denoised signal $\hat{x}^u(n)$ into clean and noise signal domain, respectively. $e_{artif}$ is a residual term and they can be formulated as follows:
\begin{eqnarray}
  x_{target}&=&\frac{x^T\hat{x}}{||x||^2}x \label{eq:t} \\
  e_{noise} &=&\frac{n^T\hat{x}}{||n||^2}n \label{eq:n}\\ 
  e_{artif} &=&\hat{x}-\left( \frac{x^T\hat{x}}{||x||^2}x + \frac{n^T\hat{x}}{||n||^2}n \right) \label{eq:a}
\end{eqnarray}
Substituting Eq.(\ref{eq:t}), (\ref{eq:n}) and (\ref{eq:a}) into Eq.(\ref{eq:sdr}), the rearranged SDR is given by
\begin{eqnarray}
\label{eq:sdr2}
\textrm{SDR}
      &=& 10\log_{10}\frac{|| \frac{x^T\hat{x}}{||x||^2}x ||^2}{||\frac{x^T\hat{x}}{||x||^2}x-\hat{x} ||^2} \nonumber \\
      &=& 10\log_{10}\frac{|| \alpha x ||^2}{|| \alpha x-\hat{x} ||^2}
 \end{eqnarray}
where $\alpha = \operatorname*{arg\,min}_\alpha ||\alpha x - \hat{x} ||$. Eq.(\ref{eq:sdr2}) coincides with SI-SDR that is another popularly used SDR definition~\cite{roux2018sdr}. In the general multiple source problems, they do not match each other. However, for single source denoising problem, we can use them interchangeably.

\begin{figure}[ht]
\vskip 0.2in
\begin{center}
\centerline{\includegraphics[width=45mm]{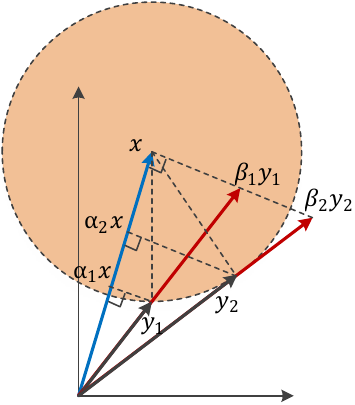}}
\caption{SDR comparison between two noisy signals with the same SNR}
\label{fig:si_sdr}
\end{center}
\vskip -0.2in
\end{figure}

SDR loss function is defined as mini-batch average of Eq.(\ref{eq:sdr2}):
\begin{equation}
\label{eq:si_sdr}
\textrm{L}_{\textrm{SDR}}=\frac{1}{B}\sum_{u=0}^{B-1} 10\log_{10}\frac{|| \alpha x^u ||^2}{|| \alpha x^u-\hat{x^u} ||^2}
\end{equation}
Compared with conventional MSE loss function, scale-invariant $\alpha$ is included as a training variable in Eq.(\ref{eq:si_sdr}). Figure~\ref{fig:si_sdr} illustrates why training $\alpha$ is important to maximize the SDR metric. For two noisy signal $y_1$ and $y_2$, they have the same SNR because they positioned at the same circle centered at the clean signal $x$. However, their SDRs are different. For $y_1$, SDR is $\frac{||\alpha_1 x ||^2}{||\alpha_1 x - y_1||^2}$, which is $\frac{||x ||^2}{||x - \beta_1 y_1||^2}$ by geometry. By the same way, SDR for $y_2$ is  $\frac{||x ||^2}{||x - \beta_2 y_2||^2}$. Clearly, $y_1$ is better than $y_2$ in terms of SDR metric but MSE criterion cannot distinguish between them because they have the same SNR. 

\subsection{PESQ Loss Function}
\label{sec:pesq}
\begin{figure}[ht]
\vskip 0.2in
\begin{center}
\centerline{\includegraphics[width=90mm]{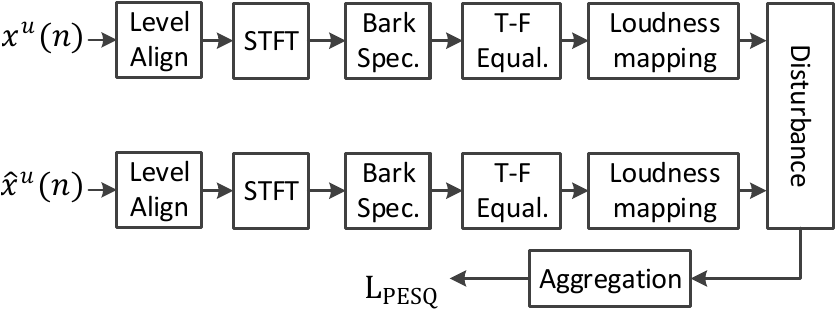}}
\caption{Block Diagram for $L_{\textrm{PESQ}}$: $x^u(n)$ is time-domain clean signal and $\hat{x}^u(n)$ is denoised time-domain signal after ISTFT.}
\label{fig:pesq}
\end{center}
\vskip -0.2in
\end{figure}
Perceptual evaluation of speech quality (PESQ)~\cite{rix2001perceptual} is ITU-T standard and it provides objective speech quality evaluation. Its value ranges from -0.5 to 4.5 and the higher PESQ means better perceptual quality. Although SDR and PESQ have high correlation, it is frequently observed that signal with smaller SDR have higher PESQ than the one with the higher SDR. For example, acoustic signal with high reverberation would have low SNR or SDR but PESQ can be much better because time-frequency equalization in PESQ can compensate the most of channel fading effects. It is just one example and there are a lot of operations that make PESQ behave much differently from SDR. Therefore, SDR loss function cannot effectively optimize PESQ. 

In this section, a new PESQ loss function is designed based on PESQ metric. Figure~\ref{fig:pesq} shows the block diagram to find PESQ loss function $\textrm{L}_{\textrm{PESQ}}$. The overall procedure is similar to the PESQ system in ~\cite{rix2001perceptual}. There are three major modifications. First, IIR filter in PESQ is removed. The reason is that time evolution of IIR filter is too deep to apply back-propagation. Second, delay adjustment routine is not necessary because training data pairs were already time-aligned. Third, bad-interval iteration was removed. PESQ improves metric calculations by detecting bad intervals of frames and updating metrics over those periods. As long as training clean and noisy data pairs are perfectly time-aligned, there's no significant impact on PESQ by removing this operation. 

\textbf{\underline{Level Alignment}}: The average power of clean and noisy speeches ranging from 300Hz to 3KHz are aligned to be $10^7$, which is a predefined power value. IIR filter gain 2.47 is also compensated in this block.

\textbf{\underline{Bark Spectrum}}: Bark spectrum block is to find the mean of linear scale frequency bins according to the Bark scale mapping. The higher frequency bins are averaged with more number of bins, which effectively provides lower weighting to them. The mapped bark spectrum power can be formulated as follows:
\begin{equation}
B^u_{c,m,i}=\frac{1}{I_{i+1}-I_i}\sum_{k=I_i}^{I_{i+1}-1}|X^u_{m,k}|^2
\end{equation}
where $I_i$ is the start of linear frequency bin number for the $i^{th}$ bark spectrun, $B^u_{c,m,i}$ is $i^{th}$ bark spectrum power of clean speech. All the positive linear frequency bins were mapped to 49 Bark spectrum bins. $B^u_{n,m,i}$ is a bark spectrum power of noisy speech and can be also found in the similar manner.

\textbf{\underline{Time-Frequency Equalization}}: Each Bark spectrum bin of clean speech is compensated by the average power ratio between clean and noisy Bark spectrum as follows:
\begin{equation}
E^u_{c,m,i}=\frac{P^u_{n,i}+c_1}{P^u_{c,i}+c_1}B^u_{c,m,i}
\end{equation}
where $P^u_{n,i}=\frac{1}{M_u}\sum_{m=0}^{M_u-1} B^u_{n,m,i}S^u_{n,m,i}$, $P^u_{c,i}=\frac{1}{M_u}\sum_{m=0}^{M_u-1} B^u_{c,m,i}S^u_{c,m,i}$, $c_1$ is a constant and $S^u_{n,m,i}$ and $S^u_{n,m,i}$ are silence masks that become 1 only when the corresponding bark spectrum power exceeds predefined thresholds. After frequency equalization, the short-term gain variation of a noisy bark spectrum is also compensated for each frame:
\begin{eqnarray}
S^u_m &=& \frac{G^u_{c,m}+c_2}{G^u_{n,m}+c_2} \label{eq:tf1}\\
S^u_m &=& 0.2S^u_{m-1}+0.8S^u_{m}\label{eq:tf2} \\
E^u_{n,m,i} &=& S^u_{m}B^u_{n,m,i} \label{eq:tf3}
\end{eqnarray}
where $G^u_{n,m}=\sum_{i=0}^{I-1}B^u_{n,m,i}$, $G^u_{c,m}=\sum_{i=0}^{I-1}E^u_{c,m,i}$ , $I$ is the size of Bark spectrum bins, and $c_2$ is a constant. 

\textbf{\underline{Loudness Mapping}}: The power densities are transformed to a Soner loudness scale using Zwicker's law~\cite{zwicker1967ohr}:
\begin{equation}
L^u_{x,m,i}=S_i\left(\frac{P_{0,i}}{0.5}\right) \left[ \left(0.5 + 0.5\frac{E^u_{x,m,i}}{P_{0,i}}\right) -1\right]
\end{equation}
where $P_{0,i}$ is the absolute hearing threshold, $S_i$ is the loudness scaling factor, and r is Zwicker power and x can be c (clean) or n (noisy).

\textbf{\underline{Disturbance Processing}}: The raw disturbance is difference between clean and noisy loudness densities with following operations:
\begin{eqnarray}
D^u_{m,i}=\max \left( L^u_{c,m,i} - L^u_{n,m,i} - \textrm{DZ}^u_{m,i},0 \right)&& \nonumber \\
          + \min \left( L^u_{c,m,i} - L^u_{n,m,i} + \textrm{DZ}^u_{m,i},0 \right)&&
\end{eqnarray} 
where $\textrm{DZ}^u_{m,i} = 0.25\min(L^u_{c,m,i},L^u_{n,m,i})$.
If absolute difference between clean and noisy loudness densities are less than 0.25 of minimum of two densities, raw disturbance becomes zero. From raw disturbance, symmetric frame disturbance is given by
\begin{equation}
\textrm{FD}^u_{m}=\sum_{i=0}^{I-1}w_i\sqrt{\frac{1}{\sum_{i=0}^{I-1}w_i}\sum_{i=0}^{I-1}\left( W_i D^u_{m,i} \right)^2}
\end{equation}
where $w_i$ is predefined weighting for bark spectrum bins.
Asymmetric frame disturbance has additional scaling and thresholding steps as follows:

\begin{eqnarray}
h^u_{m,i}&=& \left( \frac{B^u_{n,m,i}+50}{B^u_{c,m,i}+50}   \right)^{1.2} \\
h^u_{m,i}&=& \bigg\{
\begin{array}{l}
12,  \textrm{   if   } h^u_{m,i} > 12  \\
0,   \textrm{   if } h^u_{m,i} < 3
\end{array}
\end{eqnarray} 
\begin{equation}
\textrm{AFD}^u_{m} = \sum_{i=0}^{I-1}w_i\sqrt{\frac{1}{\sum_{i=0}^{I-1}w_i}\sum_{i=0}^{I-1}\left( W_i D^u_{m,i} h^u_{m,i}\right)^2}
\end{equation}

\textbf{\underline{Aggregation}}: PESQ loss function $L_{\textrm{PESQ}}$ can be found as follows:
\begin{equation}
L_{\textrm{PESQ}}=4.5-0.1d_{sym}-0.0309d_{asym}
\end{equation}
$d_{sym}$ can be found from two stage averaging:
\begin{eqnarray}
\textrm{PSQM}_s^u=\sqrt[6]{\frac{1}{20}\sum_{i=0}^{19}\left( \textrm{FD}^u_{10s+i} \right)^6} \\
d_{sym}=\sum_{u=0}^{B-1} \sqrt{\frac{1}{S^u}\sum_{s=0}^{S^u-1}\left( \textrm{PSQM}_s^u \right)^2}
\end{eqnarray}
where $S^u$ is $\lfloor \frac{M^u}{10} \rfloor$. $d_{asym}$ can also be found with similar averaging.

\subsection{Joint SDR and PESQ optimization}
\label{sec:joint}
To jointly maximize SDR and PESQ, a new loss function is defined by combining $L_{\textrm{SDR}}$ and $\textrm{L}_{\textrm{PESQ}}$:
\begin{equation}
\textrm{L}_{\textrm{SDR-PESQ}}=\textrm{L}_{\textrm{SDR}}+\alpha \textrm{L}_{\textrm{PESQ}}
\end{equation}
Another combined loss function is defined using MSE criterion instead of $\textrm{L}_{\textrm{PESQ}}$:
\begin{eqnarray}
&&\textrm{L}_{\textrm{SDR-MSE}}=\textrm{L}_{\textrm{SDR}}+ \nonumber \\
&&\alpha \sum_{u=0}^{B-1}\sum_{m=0}^{M-1}\sum_{k=0}^{K-1} \left( M^u_{m,k}|Y^u_{m,k}| -|X^u_{m,k}| \right)^2
\end{eqnarray}
By comparing $\textrm{L}_{\textrm{SDR-MSE}}$ with $\textrm{L}_{\textrm{SDR-PESQ}}$, we can evaluate PESQ improvement from the proposed PESQ loss function over MSE criterion.

%% file: results.tex
\section{Experiments and Results}
\label{sec:er}
\subsection{Experimental Settings}
\label{sec:es}

\begin{table}[h]
\caption{Train and test sets sampled from QUT-NOISE-TIMIT: The notation is X-Y-N. X refers to noise type, Y is a recorded location for the noise and N is session number.}
\label{tab:timt_qut}
\centering
\begin{tabularx}{\columnwidth}{l X}
\toprule
Train Set& {\scriptsize CAFE-FOODCOURTB-1, CAFE-FOODCOURTB-2, CAR-WINDOWNB-1, CAR-WINDOWNM-2,
  			HOME-KITCHEN-1, HOME-KITCHEN-2, REVERB-POOL1, REVERB-POOL2, 
            STREET-CITY-1, STREET-CITY2} \\
\midrule
Test Set & {\scriptsize CAFE-CAFE-1, CAR-WINUPB-1, HOME-LIVING-1, REVERB-CARPARK-1,
		   STREET-KG-1} \\
\bottomrule
\end{tabularx}
\end{table}

In order to evaluate the proposed denoising framework, we used the following three datasets:

\textbf{\underline{CHIME-4}}~\cite{vincent2017analysis}: Simulated dataset is used for training and evaluation. It is synthesized by mixing Wall Stree Journal (WSJ) corpus with noise sources from four different environments: bus, cafe, pedestrian area and street junction. Four noise conditions are applied to both train and development data. Moreover, simulated dataset has fixed SNR and therefore, it is relatively easy data to improve. 

\textbf{\underline{QUT-NOISE-TIMIT}}~\cite{dean2010qut}: QUT-NOISE-TIMIT corpus is created by mixing 5 different background noise sources with TIMIT clean speech~\cite{garofolo1993darpa}. Unlike CHIME-4, the synthesized speech sequences provide wide range of SNR categories from -10dB to 15dB SNR. For training set, only -5 and 5 dB SNR data were used but test set contains all SNR ranges. The noise and location information used for train and test sets are summarized in Table~\ref{tab:timt_qut}.  

\textbf{\underline{VoiceBank-DEMAND}}~\cite{valentini2016investigating}: 30 speakers selected from Voice Bank corpus~\cite{veaux2013voice} were mixed with 10 noise types: 8 from Demand dataset~\cite{thiemann2013diverse} and 2 artificially generated one. Test set is generated with 5 noise types from Demand that does not coincide with those for training data. VoiceBank-DEMAND corpus was used to evaluate generative models such as SEGAN~\cite{pascual2017segan}, TF-GAN~\cite{soni2018time} and WAVENET~\cite{rethage2018wavenet}.

\subsection{Model Comparison}

\begin{table}[ht]
\caption{SDR and PESQ results between different denoising architectures on CHIME-4 corpus. The result for DNN* came from~\cite{bando2018statistical}.}
\label{tab:model}
\vskip 0.1in
\centering
\begin{tabularx}{\columnwidth}{l X X r r}
\toprule
Model 		& G-L Iter.  	& Loss Type	& SDR 		& PESQ  \\
\midrule
Noisy Input	& -				&-		& 5.80 	&	1.267 \\
OM-LSA		&	-			&-		&	8.50 & 1.514 \\
DNN*		&	-			&IRM	&	10.93 &	- \\
CNN DAE		&	1			&IAM	&   11.30 &	1.507 \\
CNN DAE		&	10			&IAM	&	11.21 &	1.491	\\
CNN-BLSTM	&	1			&IAM	&	11.91 &	1.822 \\
CNN-BLSTM	&	10			&IAM	&	12.06 &	1.829 \\
\bottomrule
\end{tabularx}
\end{table}

\begin{table*}[ht]
\caption{SDR and PESQ results on QUT-NOISE-TIMIT: Test set consists of 6 SNR ranges:-10, -5, 0, 5, 10, 15 dB. The highest SDR or PESQ scores for each SNR test data were highlighted with bold fonts.}
\label{tab:tq}
\vskip 0.1in
\centering
\begin{tabular}{l r r r r r r | r r r r r r }
\toprule
	& \multicolumn{6}{c|}{SDR}	& \multicolumn{6}{c}{PESQ} \\
Loss Type	& -10 dB & -5 dB & 0 dB & 5 dB & 10 db & 15 dB & -10 dB & -5 dB & 0 dB & 5 dB & 10 db & 15 dB\\
\midrule
Noisy Input & -11.82 & -7.33 & -3.27 & 0.21 & 2.55 & 5.03 & 1.07 & 1.08 & 1.13 & 1.26 & 1.44 & 1.72 \\
IAM & -3.23  & 0.49 & 2.79 & 4.63 & 5.74 & 7.52 & 1.29 & 1.47 & 1.66 & 1.88 & 2.07 & 2.30 \\ 
PSM & -2.95 & 0.92 & 3.37 & 5.40 &6.64 & 8.50 & 1.30 & 1.49 & 1.71 & 1.94 & 2.15 & 2.37 \\
SNR & -2.79 & 1.36 & 3.70 & 5.68 & 6.18 & 8.44 & 1.29 & 1.46 & 1.68 & 1.93 & 2.14 & 2.38 \\
SDR & -2.66 & 1.55 & 4.13 & 6.25 & 7.53 & 9.39 & 1.26 & 1.42 & 1.65 & 1.92 & 2.16 & 2.41  \\
SDR-MSE & -2.53 & 1.57 & 4.10 & 6.31 & 7.60 & 9.43 & 1.29 & 1.47 & 1.68 & 1.93 & 2.15 & 2.39 \\
SDR-PESQ & \textbf{-2.31} & \textbf{1.80} & \textbf{4.36} & \textbf{6.51} & \textbf{7.79} & \textbf{9.65} & \textbf{1.43} & \textbf{1.65} & \textbf{1.89} & \textbf{2.16} & \textbf{2.35} & \textbf{2.54}   \\  
\bottomrule
\end{tabular}
\end{table*}

Table~\ref{tab:model} showed SDR and PESQ performance for various denoising networks based on CHIME-4 corpus. OM-LSA is a baseline statistical scheme that presented 2.7 dB SDR gain and 0.25 PESQ improvement. For neural network-based approaches, the author in~\cite{bando2018statistical} presented 10.93 dB SDR for 5-layer DNN. In this paper, two models were trained: CNN-based denoising autoencoder (DAE) and CNN-BLSTM. The generator architecture in~\cite{pascual2017segan} was used for CNN-DAE. It presented SDR improvement over DNN model but PESQ performance is worse than OM-LSA. For CNN-BLSTM, both SDR and PESQ significantly improved, which is why we chose it as our model architecture.

Iterative Griffin-Lim algorithm~\cite{griffin1984signal} was evaluated for CNN-DAE and CNN-BLSTM at the inference time. SDR and PESQ for two networks converged after 10 iterations. CNN-BLSTM showed 0.15 dB SDR gain from 10 iterations, on the other hand, CNN DAE showed 0.1 dB loss. Although iterative Griffin-Lim algorithm reduces amplitude spectrum mismatch for each iteration, phase spectrum also changes from update. Phase spectrum update is not predictable. As shown in this experiment, sometimes, it is beneficial but in other case, it is not. Due to unstable SDR performance, iterative Griffin-Lim algorithm was not used in the inference time. Iterative Griffin-Lim in the training stage would be evaluated at the later section.     

\subsection{Main Results}

Table~\ref{tab:tq} compared SDR and PESQ performance between different denoising methods on QUT-NOISE-TIMIT corpus. All the schemes were based on the same CNN-BLSTM model and trained with -5 and +5 dB SNR data as explained at Section~\ref{sec:es}. IAM and PSM are two spectrum mask estimation schemes explained at Section~\ref{sec:sme}. SDR refers $\textrm{L}_{\textrm{SDR}}$ at Section~\ref{sec:si-sdr} and SDR-MSE and SDR-PESQ correspond to $\textrm{L}_{\textrm{SDR-PESQ}}$ and $\textrm{L}_{\textrm{SDR-MSE}}$ at Section~\ref{sec:joint}, respectively.

The proposed end-to-end scheme based on $\textrm{L}_{\textrm{SDR}}$ showed significant SDR gain over spectrum mask schemes for all SNR ranges. However, for PESQ, it did not show similar improvement due to metric mismatch. For example, PESQ performance degraded on the low SNR ranges such as -10, -5 and 0 dB over PSM. 

Two proposed joint optimization schemes were evaluated. First, $\textrm{L}_{\textrm{SDR-MSE}}$ showed improvement on both SDR and PESQ metrics for the most of SNR ranges. However, the improvement is marginal and it still suffered from PESQ loss on low SNR ranges over PSM.
Second, $\textrm{L}_{\textrm{SDR-PESQ}}$ loss function improved both SDR and PESQ metrics with large margin over all other schemes. By combining $\textrm{L}_{\textrm{PESQ}}$ loss function with $\textrm{L}_{\textrm{SDR}}$, both SDR and PESQ performances were significantly improved.

At Section~\ref{sec:si-sdr}, we claimed MSE loss function cannot correctly maximize SDR. To evaluate this statement, we trained CNN-BLSTM with MSE criterion, which is the SNR entry in Table~\ref{tab:timt_qut}. Compared with $\textrm{L}_{\textrm{SDR}}$, SDR performance for SNR loss function degraded for all SNR ranges. One thing to note is that PESQ performance does not degrade. Therefore, this also showed that $\textrm{L}_{\textrm{SDR}}$ or SNR loss functions couldn't correctly impact PESQ metric. 
\begin{table}[ht]
\caption{SDR and PESQ results on CHIME-4}
\label{tab:c4}
\vskip 0.1in
\centering
\begin{sc}
\begin{tabular}{l r r}
\toprule
Loss Type		& SDR 		& PESQ  \\
\midrule
IAM				&	11.91	&	1.822 \\
PSM				&	12.08	&	1.857 \\
SDR			&	12.43	&	1.699 \\
SDR-MSE		&	12.44	&	1.758 \\
SDR-PESQ		&	\textbf{12.59}	&	\textbf{1.953} \\
\bottomrule
\end{tabular}
\end{sc}
\vskip -0.1in
\end{table}

Table~\ref{tab:c4} showed SDR and PESQ result on CHIME-4. The result is similar to QUT-NOISE-TIMIT. $\textrm{L}_{\textrm{SDR}}$ presented significant SDR improvement but PESQ performance degraded. The proposed joint SDR and PESQ optimization presented the best performance both on SDR and PESQ metrics.

\begin{figure}[ht]
\vskip 0.1in
\begin{center}
\centerline{\includegraphics[width=90mm]{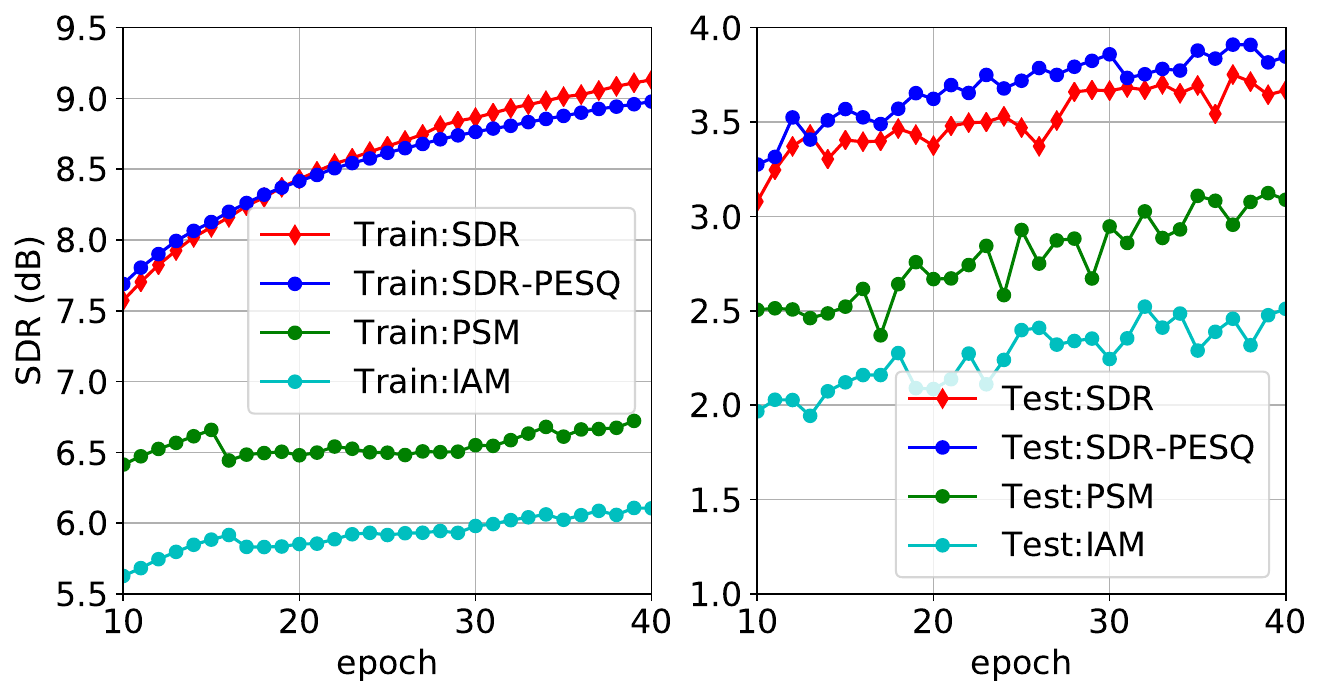}}
\caption{Train and test SDR curve comparison between different denoising schemes}
\label{fig:cv}
\end{center}
\vskip -0.2in
\end{figure}

The proposed joint optimization loss, $\textrm{L}_{\textrm{SDR-PESQ}}$ improved both SDR and PESQ even compared with $\textrm{L}_{\textrm{SDR}}$. It is logical to guess $\textrm{L}_{\textrm{SDR}}$ should present the best SDR performance. However, for both CHIME-4 and QUT-NOISE-TIMIT corpora, PESQ loss function in  $\textrm{L}_{\textrm{SDR-PESQ}}$ was also helpful to improve SDR metric. The reason can be explained by train and test curves in Figure~\ref{fig:cv}. The training SDR curve for $\textrm{L}_{\textrm{SDR}}$ reached the highest value as expected but it showed the lower SDR on the test set than $\textrm{L}_{\textrm{SDR-PESQ}}$. Clearly, $\textrm{L}_{\textrm{PESQ}}$ also acted as a regularizer to avoid overffiting. We tried other regularization terms such as $L_1$ and $L_2$ norms but $\textrm{L}_{\textrm{PESQ}}$  was more effective to improve generalization to unseen data than other general regularization methods.       
\subsection{Comparison with Generative Models}
\begin{table}[ht]
\caption{Evaluation on VoiceBank-DEMAND corpus: DCUnet-10 and DCUnet-20 are based on real-valued network (RMRn).}
\label{tab:ge}
\vskip 0.1in
\centering
\begin{tabularx}{\columnwidth}{l X X X X r}
\toprule
Models	& CSIG	& CBAK	& COVL	& PESQ	& SSNR \\
\midrule
Noisy Input	&	3.37	&	2.49	&	2.66	&	1.99	&	2.17 \\
SEGAN		&	3.48	&	2.94	&	2.80	&	2.16	&	7.73 \\
WAVENET		&	3.62	&	3.23	&	2.98	&	-		&	-	\\
TF-GAN		&	3.80	&	3.12	&	3.14	&	2.53	&	- \\
DCUnet-10	&	3.70	&	3.22	&	3.10	&	2.52	&	9.40 \\
DCUnet-20	&	\textbf{4.12}	&	3.47	&	3.51	&	2.87	&	9.96 \\
WSDR		&	3.54	&	3.24	&	3.01	&	2.51	&	9.93	\\
SDR-PESQ	&	4.09	&	\textbf{3.54}	&	\textbf{3.55}	&	\textbf{3.01}	&	\textbf{10.44} \\
\bottomrule
\end{tabularx}
\end{table}
Table~\ref{tab:ge} showed comparison with other generative models. All the results except our end-to-end model came from the original papers: SEGAN~\cite{pascual2017segan}, WAVENET~\cite{rethage2018wavenet} and TF-GAN~\cite{soni2018time}. CSIG, CBAK and COVL are objective measures where high value means better quality of speech~\cite{hu2008evaluation}. CSIG is mean opinion score (MOS) of signal distortion, CBAK is MOS of background noise intrusiveness and COVL is MOS of the overall effect. SSNR is Segmental SNR defined in~\cite{quackenbush1986objective}.

The proposed SDR and PESQ joint optimization scheme outperformed all the generative models in all objective measures. Unfortunately, the original papers didn't provide SDR metric. SSNR metric is similar to SDR but basically, SSNR is scale-sensitive metric and therefore, SDR loss function would not be optimal to maximize SSNR. Nevertheless, SSNR performance for the SDR-PESQ loss function showed significant gain over other generative models. 

The results for DCUnet-10 and DCUnet-20 in Table~\ref{tab:ge} came from the recent paper~\cite{choi2018phase}. This paper suggested a couple of fancy ideas. One of main contributions was phase compensation scheme, which showed significant improvement on VoiceBank-DEMAND corpus. Our proposed joint optimization scheme was based on amplitude spectrum estimation. However, it is not limited to real mask estimation but can also enjoy phase compensation gain if applied. Therefore, the gain from complex mask should be removed for fair comparison. Fortunately, this paper provided its scheme based on real-valued network (RMRn), which were shown in Table~\ref{tab:ge}. Our joint optimization scheme, $\textrm{L}_{\textrm{SDR-PESQ}}$ presented better performance on almost all the objective measures. In order to remove performance variation from model difference, a CNN-BLSTM model was also trained with the weighted SDR loss function proposed in this paper, which is WSDR in Table~\ref{tab:ge}. It showed comparable performance to DCUnet-10. Compared with $\textrm{L}_{\textrm{SDR-PESQ}}$, it showed large loss especially for PESQ and SSNR.   

\subsection{Training with Iterative Griffin-Lim Algorithm}
\begin{table}[ht]
\caption{SDR and PESQ evaluation of Iterative Griffin-Lim algorithm applied to joint SDR and PESQ optimization}
\label{tab:gr}
\vskip 0.1in
\centering
\begin{tabular}{ccc}
\toprule
Griffin-Lim Iteration		& SDR  		& PESQ  \\
\midrule
1		&	\textbf{12.59}	&	\textbf{1.953} \\
2		&	11.59				&	1.679 \\
3		&	12.35				&	1.949 \\
4		&	11.61				&	1.638 \\
\bottomrule
\end{tabular}
\vskip -0.1in
\end{table}

K-MISI scheme for blind source separation~\cite{wang2018end} applied multiple STFT-ISTFT operations similar to iterative Griffin-Lim algorithm. It showed SDR improvement by increasing the number of iterations. We also applied iterative Griffin-Lim algorithm to our end-to-end joint optimization at the training stage. Table~\ref{tab:gr} showed SDR and PESQ performance for each Griffin-Lim iteration. Multiple iteration of Griffin-Lim algorithm only hurt SDR and PESQ performance unlike K-MISI. The key difference in K-MISI is that the reconstructed time-domain signal is redistributed between multiple sources for each iteration, which could provide substantial enhancement of source separation. For single source denoising problem, single iteration of Griffin-Lim algorithm presented the best performance.

%% file: conclusion.tex
\section{Conclusion}
\label{sec:conclusion}
In this paper, a new end-to-end multi-task denoising scheme was proposed. The proposed scheme resolved two issues addressed before: Spectrum and metric mismatches. First, two metric-based loss functions are defined: SDR and PESQ loss functions. Second, two newly defined loss functions are combined for joint SDR and PESQ optimization. Finally, the combined loss function is optimized based on the reconstructed time-domain signal after Griffin-Lim ISTFT in order to avoid spectrum mismatch. The experimental result presented that the proposed joint optimization scheme significantly improved SDR and PESQ performances over both spectrum mask estimation schemes and generative models.